# Elastic properties and chemical bonding in ternary arsenide $SrFe_2As_2$ and quaternary oxyarsenide LaFeAsO - basic phases for new 38-55K superconductors

I.R. Shein,* and A.L. Ivanovskii

*Institute of Solid State Chemistry, Ural Branch of the Russian Academy of Sciences, 620041, Ekaterinburg, Russia*

**We report the first-principle FLAPW-GGA calculations of the elastic properties of two related layered phases, namely, the ternary arsenide $SrFe_2As_2$ and the quaternary oxyarsenide LaFeAsO - basic phases for the newly discovered "122" and "1111" 38-55K superconductors. The independent elastic constants ($C_{ij}$), bulk moduli, compressibility, and shear moduli are evaluated and discussed. The numerical estimates of the elastic parameters of the polycrystalline $SrFe_2As_2$ and LaFeAsO ceramics are performed for the first time. Additionally, the peculiarities of chemical bonding in these phases are discussed.**

## 1. Introduction

Recently, two new groups of FeAs-based high $T_C$ superconductors (SC's) have been discovered [1,2]. For one group (the so-called "1111" type *Ln*FeAsO, where *Ln* are light rare-earth metals La, Ce….Gd, Tb, Dy), the critical temperature may reach 55K as a result of hole or electron doping [1,3-7]. More recently, ternary oxygen-free arsenides termed as "122" systems, namely, $AFe_2As_2$, where *A* are alkaline earth metals Ca, Sr or Ba, were discovered [2] as parent phases for new hole-doped 20-38K SC's [8-10]. All of these FeAs-based SCs (i) are derivatives of magnetic spin density wave (SDW) systems; (ii) adopt layered tetragonal structures, where $[LnO]^{1+}$ layers (or planar sheets of ions $A^{2+}$) are sandwiched between $[FeAs]^{1-}$ layers formed by edge-shared tetrahedra $FeAs_4$ and (iii) reveal a two-dimensional electronic structure. It is believed that superconductivity emerges in the [FeAs] layers while the [*Ln*O] layers (or planar sheets of alkaline earth atoms) provide an electron (hole) reservoir in doping procedures [1-10].

* Corresponding author.
E-mail address: shein@ihim.uran.ru

These findings generated a tremendous interest in the scientific community and triggered much activity in research of unconventional superconducting mechanisms for these FeAs-based systems. Additionally, some advanced applications for these materials were proposed. For example, the upper critical field $H_{c2}(0)$ for LaFeAsO$_{1-x}$F$_x$ can exceed 60 - 65 T [11], making these materials potentially useful in very high magnetic field applications, see also [12]. Other possible applications of these materials may be found in thermoelectric cooling modules in the liquid nitrogen temperature range [13].

Thus, the mechanical properties of FeAs-based SC's are of great importance for their material science in view of future technological applications. Besides, possible correlations between the critical temperature $T_C$ and the mechanical parameters are of interest. For example, it was supposed [14] that high $T_C$'s are associated with low values of the bulk modulus $B$, i.e. with high compressibility $\beta$. Really, many SC's with enhanced critical temperatures (such as YBCO, MgB$_2$, MgCNi$_3$, borocarbides $Ln$M$_2$B$_2$C, carbide halides of rare earth metals, $Ln_2$C$_2$X$_2$ *etc*) are relatively soft: their bulk moduli do not exceed $B \leq 200$ GPa ($\beta \geq 0.005$ 1/GPa) [15-18]. On the other hand, the superconducting transition (to $T_C \sim 11$K) was found for such a hard and incompressible material as boron - doped diamond [19].

In the present study, we report a comparative *ab initio* analysis of the elastic properties of the ternary arsenide SrFe$_2$As$_2$ and the quaternary oxyarsenide LaFeAsO (as representative species for the two above mentioned "122" and "1111" types of parent phases of the new FeAs-based 38-55K superconductors) which have not yet been investigated either experimentally or theoretically.

As a result, their independent elastic constants ($C_{ij}$), bulk moduli, compressibility, and shear moduli were evaluated and discussed, and numerical estimates of the elastic parameters for the polycrystalline SrFe$_2$As$_2$ and LaFeAsO ceramics were performed for the first time. Additionally, the peculiarities of chemical bonding for SrFe$_2$As$_2$ as compared with LaFeAsO were discussed.

## 2. Computational details

Our calculations were carried out by means of the full-potential method with mixed basis APW+lo (LAPW) implemented in the WIEN2k suite of programs [20]. The generalized gradient approximation (GGA) to exchange-correlation potential in the PBE form [21] was used. The plane-wave expansion was taken to $R_{MT} \times K_{MAX}$ equal to 7, and the $k$ sampling with $10\times10\times4$ $k$-points in the Brillouin zone was used. La ($5s^25p^65d^16s^2$), O ($2s^22p^4$), Fe ($3d^64s^2$), As ($4p^34s^2$) and Sr ($5s^25p^0$) were treated as valence states.

The calculations were performed with full-lattice optimizations including internal parameters $z_{La}$ and $z_{As}$. The self-consistent calculations were considered to be converged when the difference in the total energy of the crystal did not exceed 0.01 mRy as calculated at consecutive steps, and the nonmagnetic state for SrFe$_2$As$_2$ and LaFeAsO was treated.

The hybridization effects were analyzed using the densities of states (DOS), which were obtained by a modified tetrahedron method [22]. To quantify the amount of electrons redistributed between the atomic sublattices and the adjacent layers (sheets), *i.e.* for the discussion of the ionic bonding, we performed also a Bader [23] analysis.

## 3. Results and discussion
### 3.1. Elastic properties

Firstly, six independent elastic constants ($C_{ij}$; namely $C_{11}$, $C_{12}$, $C_{13}$, $C_{33}$, $C_{44}$ and $C_{66}$) for the tetragonal SrFe$_2$As$_2$ and LaFeAsO phases were evaluated by calculating the stress tensors on different deformations applied to the equilibrium lattice of the tetragonal unit cell, whereupon the dependence between the resulting energy change and the deformation was determined, Table 1. For LaFeAsO, all these elastic constants were positive and satisfied the well-known Born's criteria for tetragonal crystals: $C_{11} > 0$, $C_{33} > 0$, $C_{44} > 0$, $C_{66} > 0$, $(C_{11} - C_{12}) > 0$, $(C_{11} + C_{33} - 2C_{13}) > 0$ and $\{2(C_{11} + C_{12}) + C_{33} + 4C_{13}\} > 0$. However, for SrFe$_2$As$_2$ our calculations showed $C_{44} \sim 0$. This implies that this phase lies on the border of mechanical stability.

Secondly, the calculated elastic constants allowed us to obtain the macroscopic mechanical parameters for SrFe$_2$As$_2$ and LaFeAsO phases, such as bulk moduli ($B$) and shear moduli ($G$) – for example, using the Voigt (V) [24] approximation, as:

$$B_V = 1/9\{2(C_{11} + C_{12}) + C_{33} + 4C_{13}\};$$
$$G_V = 1/30(M + 3C_{11} - 3C_{12} + 12C_{44} + 6C_{66});$$

where $C^2 = (C_{11} + C_{12})C_{33} - 2C_{13}^2$ and $M = C_{11} + C_{12} + 2C_{33} - 4C_{13}$. The results obtained are presented in Table 1. From these data we see that for both SrFe$_2$As$_2$ and LaFeAsO phases $B_V > G_V$; this means that a parameter limiting the mechanical stability of these materials is the shear modulus $G_V$. That is obvious enough due to the layered structure and the sharply anisotropic bonding picture for these phases, as is clearly seen on the electron density maps, Fig. 1, see also below.

Next, as SrFe$_2$As$_2$ and LaFeAsO species are usually prepared and investigated as polycrystalline ceramics (see [1-12]) in the form of aggregated mixtures of micro-crystallites with a random orientation, it is useful to estimate the corresponding parameters for these polycrystalline materials from the elastic constants of the single crystals.

For this purpose we also calculated monocrystalline bulk moduli ($B$) and shear moduli ($G$) in Reuss approximation (R: $B_R$ and $G_R$) [25], and then we utilized the Voigt-Reuss-Hill (VRH) approximation. In this approach, according to Hill [26], the Voigt and Reuss averages are limits and the actual effective moduli for polycrystals could be approximated by the arithmetic mean of these two limits. Then, one can calculate the averaged compressibility ($\beta_{VRH} = 1/B_{VRH}$), Young moduli ($Y_{VRH}$), and from $B_{VRH}$, $G_{VRH}$ and $Y_{VRH}$ it is possible to evaluate the Poisson's ratio ($v$). All these parameters are listed in Table 2. Certainly, all these estimations were performed in the limit of zero porosity of SrFe$_2$As$_2$ and LaFeAsO ceramics.

From our results we see that the bulk moduli for SrFe$_2$As$_2$ and LaFeAsO ceramics are rather small (< 100 GPa) and are less than, for example, the bulk moduli for other known superconducting species such as MgB$_2$, MgCNi$_3$, YBCO and YNi$_2$B$_2$C for which $B$ vary from 115 to 200 GPa [15-18]. Thus, as compared with other superconductors, SrFe$_2$As$_2$ and LaFeAsO are soft materials.

In turn, comparing SrFe$_2$As$_2$ and LaFeAsO, we see that $B_{VRH}$(LaFeAsO) > $B_{VRH}$(SrFe$_2$As$_2$) and $G_{VRH}$(LaFeAsO) > $G_{VRH}$(SrFe$_2$As$_2$). Accordingly, the compressibility of these species changes as $\beta$(SrFe$_2$As$_2$) < $\beta$(LaFeAsO). The Young modulus $Y_{VRH}$ has also the minimal and maximal values for SrFe$_2$As$_2$ and LaFeAsO ceramics, respectively. The most simple explanation of these data follows from the known correlation between the bulk moduli and cell volumes ($B \sim V_o^{-k}$ [27]). In our case, the optimized cell volumes are $V_o(SrFe_2As_2)$ = 183.6 Å$^3$ > $V_o(LaFeAsO)$ = 141.2 Å$^3$.

Finally, according to the criterion [28], a material is brittle if the *B/G* ratio is less than 1.75. In our case, for $SrFe_2As_2$ and LaFeAsO these values are 26.83 and 1.74, respectively. This means that LaFeAsO lies on the border of brittle behavior, whereas $SrFe_2As_2$ will behave as a very brittle material. Besides, the values of the Poisson ratio (*v*) are minimal for covalent materials (*v* = 0.1) and grow essentially for ionic species [29]. In our case, the values of *v* for LaFeAsO and $SrFe_2As_2$ are about 0.26 and 0.48, respectively, *i.e.* a considerable ionic contribution in intra-atomic bonding takes place for these phases, see Fig. 1.

*3.2. Chemical bonding*

Let us discuss the comparative peculiarities of the chemical bonding for $SrFe_2As_2$ and LaFeAsO phases. Note that up to now the bonding scenario was discussed only for some "1111" phases, see [30].

The total and site projected DOS for $SrFe_2As_2$ and LaFeAsO as calculated for equilibrium geometries are given in Fig. 2. The valence band (VB) for LaFeAsO includes two main subbands, where the high-energy subband (ranges from -5.5 to -2.2 eV below the Fermi level, $E_F$) comprises the contributions of Fe 3*d*, As 4*p* states (from [FeAs] layers) and O 2*p* and La *p* states (from [LaO] layers). The topmost region of the VB (ranges from -2.2 eV to $E_F$) is derived basically from Fe 3*d* states. Thus, we can conclude that (i) the electronic bands around the Fermi level are formed mainly by Fe 3*d* states of [Fe-As] layers, and these delocalized states are responsible for metallic-like Fe-Fe bonds; (ii) the general bonding mechanism in LaFeAsO does not correspond to the "pure" ionic picture, rather it includes covalent interactions inside [LaO] and [FeAs] layers, as well as inter-layer [LaO]-[FeAs] interactions due to hybridization of As 4*p* - La *p* orbitals.

For $SrFe_2As_2$, the VB bottom (in the region from -5.8 eV to -2.2 eV) is formed of comparable contributions from the As 4*p* and Fe 3*d* states, while the near-Fermi region contains the main contributions from the Fe 3*d* states. Besides, it is noteworthy that the contributions from the valence states of Sr to the VB are negligible. This means that the Sr sheets and the [FeAs] layers are linked exclusively by ionic interactions, as distinct from the quaternary oxyarsenide LaFeAsO.

The formation of the mentioned covalent bonds for $SrFe_2As_2$ and LaFeAsO is also evident from the charge density pictures (Fig. 1) where it is clearly shown that the charge distributions are not spherically symmetric but are strongly deformed along Fe-As and La-O bond directions inside the corresponding layers.

To analyze the ionic contributions to chemical bonding, we performed a Bader [23] analysis. In this approach, each atom of a crystal is surrounded by an effective surface that runs through minima of the charge density, and the total charge of an atom (the so-called Bader charge, $Q^B$) is determined by integration within this region. The calculated atomic $Q^B$ as well as the corresponding charges as obtained from a purely ionic model ($Q^i$) and their differences ($\Delta Q = Q^B - Q^i$) are presented in Table 3. The obtained results show that the charge transfer from [LaO] to [FeAs] layers is about 0.40 and from Sr atomic sheets to [FeAs] layers is about 0.33 electrons per formula unit for LaFeAsO and $SrFe_2As_2$ respectively. Naturally, taking into account the metallic-like type of the both compounds, these values are much smaller than those assumed for the idealized ionic picture. Let us note also that the charge transfer between atoms inside each layer (for example, for LaFeAsO: from Fe to As at about 1.7 e, from La to oxygen at about 1.1 e *etc*, see Table 3) is much higher than between the adjacent layers. This implies that individual ionic bonds between various atoms and layers are highly anisotropic.

Thus, the presented results allow us to draw the following common conclusions. For both $SrFe_2As_2$ and LaFeAsO phases, the chemical bonding is of a complex and anisotropic character which can be explained by the contributions of separate atomic sublattices to the common bonding system. Namely, for LaFeAsO:

- the Fe atoms participate in Fe-As covalent bonds (due to Fe $3d$ - As $4p$ hybridization), Fe-Fe metallic-like bonds (due to delocalized Fe $3d$ states), and in Fe-As ionic bonds (due to Fe $\rightarrow$ As charge transfer);
- the As atoms participate in As-Fe covalent bonds (due to Fe $3d$ - As $4p$ hybridization), As - Fe ionic bonds (due to Fe $\rightarrow$ As charge transfer), and in weak covalent "intra-layer" As-La bonds (due to As $p$ – La $p$ hybridization);
- the La atoms participate in La-O covalent bonds (due to La $p$ - O $2p$ hybridization), La - O ionic bonds (due to La $\rightarrow$ O charge transfer), and in weak covalent "inter-layer" La-As bonds (due to La $p$ – As $p$ hybridization);
- the O atoms participate in O-La covalent bonds (due to O $2p$- La $p$ hybridization), and O - La ionic bonds (due to La $\rightarrow$ O charge transfer); and finally,
- all of the mentioned atoms: Fe and As (in [FeAs] layers) and La and O (in [LaO] layers) participate in "inter-layer" ionic bonds (due to [LaO] $\rightarrow$ [FeAs] charge transfer).

For $SrFe_2As_2$, the bonding picture has some important features. Though the main bond types inside the [FeAs] layers (i.e. covalent, ionic Fe-As and metallic-like Fe-Fe bonds) are preserved, the interactions between the adjacent layers [FeAs] and atomic sheets [*A*] are exclusively of ionic type. As a result, the inter-layer bonds (which are responsible for the common stability of crystals) are weaker for $SrFe_2As_2$ than for LaFeAsO: for $SrFe_2As_2$ the inter-layer [FeAs]-[Sr] covalent bonds are absent, and the charge transfer [LaO] $\rightarrow$ [FeAs] is about 20% smaller than for LaFeAsO, see Table 3. Thus, these results may qualitatively explain why $SrFe_2As_2$ is a softer material as compared with LaFeAsO, as was established from our estimations of the elastic parameters.

Let us also note that these phases may be described as *quasi-two-dimensional ionic metals* where the conduction is strongly anisotropic, happening in [FeAs] layers, whereas they are stabilized mainly by electron transfer between the adjacent layers (or between the layers and atomic sheets, Fig. 1), *i.e.* due to ionic interactions, see also [31].

**4. Summary**

In conclusion, the first-principles FLAPW-GGA total energy calculations were performed to predict the elastic properties for mono- and polycrystalline $SrFe_2As_2$ and LaFeAsO as basic phases for the newly discovered 38-55K SCs. Our analysis showed that LaFeAsO is a mechanically stable anisotropic material, whereas $SrFe_2As_2$ lies on the border of mechanical stability. The parameter limiting their mechanical stability is the shear modulus. Both phases are relatively soft materials ($B < 100$ GPa) with high compressibility and will exhibit a brittle behavior; $SrFe_2As_2$ being softer than LaFeAsO. The latter inference may be qualitatively explained on the basis of comparative analysis of the chemical bonding pictures for the both species.

**Table 1.** Calculated elastic constants ($C_{ij}$, in GPa), bulk moduli ($B$, in GPa) and shear moduli ($G$, in GPa) for tetragonal monocrystalline $SrFe_2As_2$ and LaFeAsO.

| phase / parameters | $SrFe_2As_2$ | LaFeAsO |
|---|---|---|
| $C_{11}$ | 166.1 | 191.9 |
| $C_{12}$ | 30.2 | 55.9 |
| $C_{13}$ | 36.9 | 61.6 |
| $C_{33}$ | 65.0 | 144.8 |
| $C_{44}$ | ~0 | 44.1 |
| $C_{66}$ | 80.5 | 77.9 |
| $B_V$ * | 67.3 | 98.5 |
| $G_V$ * | 32.1 | 56.5 |

* in Voigt approximations

**Table 2**. Calculated values of some elastic parameters for polycrystalline $SrFe_2As_2$ and LaFeAsO ceramics as obtained in the Voigt-Reuss-Hill approximation: bulk moduli ($B_{VRH}$, in GPa), compressibility ($\beta_{VRH}$, in GPa$^{-1}$), shear moduli ($G_{VRH}$, in GPa), Young moduli ($Y_{VRH}$, in GPa) and Poisson's ratio ($v$).

| oxypnictide | $B_{VRH}$ | $\beta_{VRH}$ | $G_{VRH}$ | $Y_{VRH}$ | $v$ |
|---|---|---|---|---|---|
| $SrFe_2As_2$ | 61.7 | 0.01621 | 2.3 | 6.8 | 0.482 |
| LaFeAsO | 97.9 | 0.01022 | 56.2 | 141.5 | 0.259 |

**Table 3.** Atomic charges (in e) for $SrFe_2As_2$ and LaFeAsO and for [FeAs], [LaO] layers and Sr sheets as obtained from a purely ionic model ($Q^i$), from Bader analysis ($Q^B$) and their differences ($\Delta Q = Q^B - Q^i$).

| phases | charges | La(Sr) | Fe | As | O | [LaO] (0.5 Sr) | [FeAs] |
|---|---|---|---|---|---|---|---|
| LaOFeAs | $Q^i$ | 8(+3) | 6(+2) | 8(-3) | 8(-2) | - | - |
|  | $Q^B$ | 9.115 | 7.722 | 5.887 | 7.276 | - | - |
|  | $\Delta Q$ | 1.115 | 1.722 | -2.113 | -0.724 | 0.391 | -0.391 |
| $SrFe_2As_2$ | $Q^i$ | 8(+2) | 6(+2) | 8(-3) | - | - | - |
|  | $Q^B$ | 8.657 | 7.912 | 5.759 | - | - | - |
|  | $\Delta Q$ | 0.657 | 1.912 | -2.241 | - | 0.329 | -0.329 |

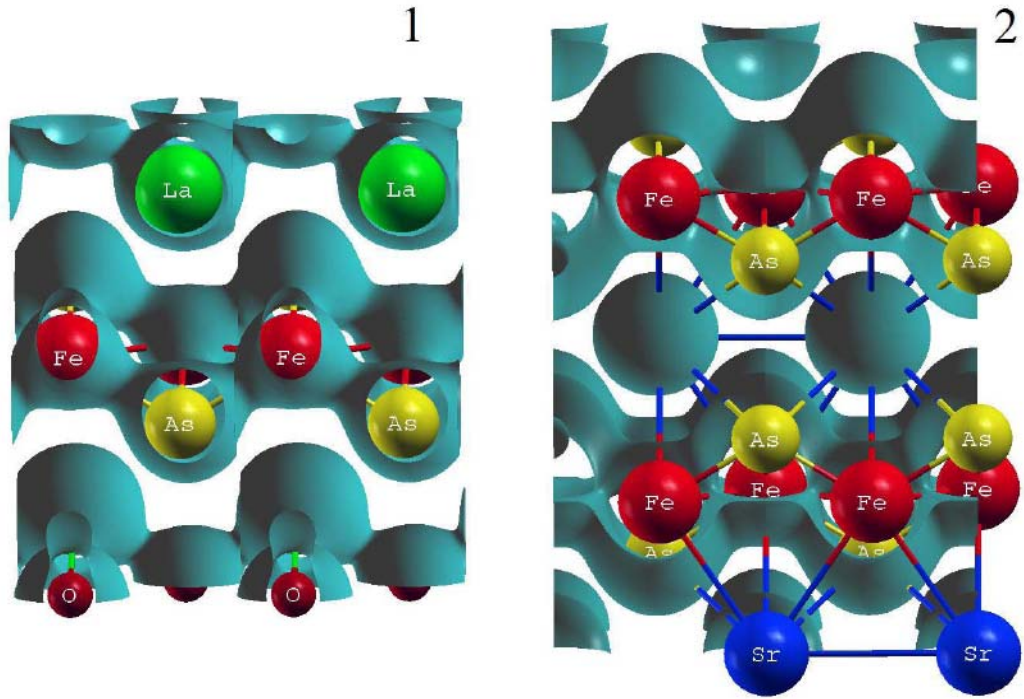

**Figure 1.** (*Color online*) Isoelectronic ($\rho = 0.36$ e/Å$^3$) surfaces for 1- LaFeAsO and 2 - SrFe$_2$As$_2$ according to FLAPW-GGA calculations.

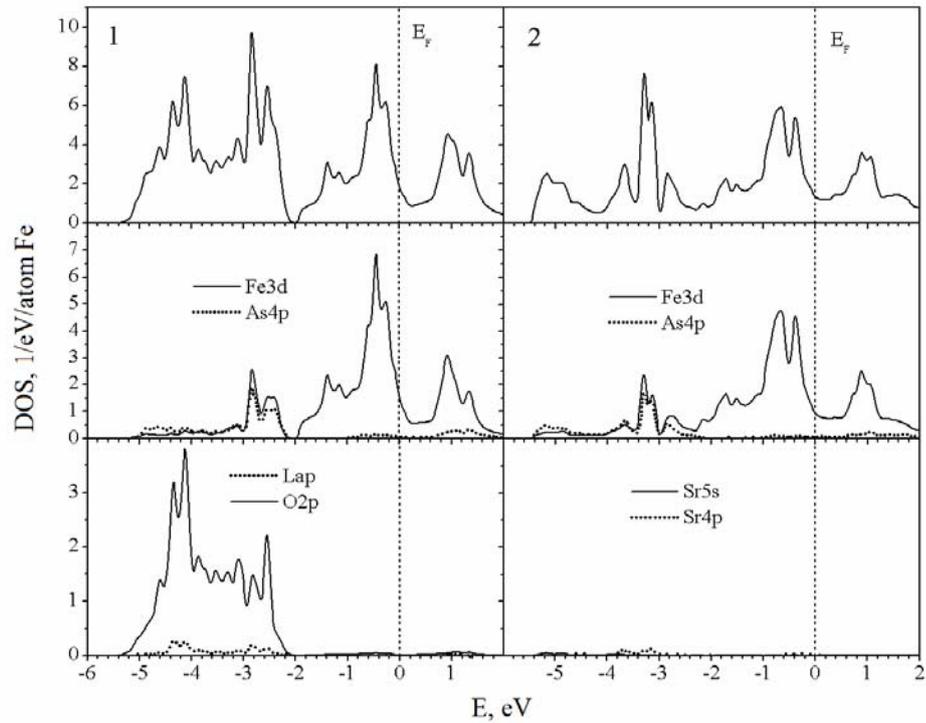

**Figure 2.** Total (upper panel) and partial densities of states for 1- LaFeAsO and 2 - SrFe$_2$As$_2$ according to FLAPW-GGA calculations.